\documentclass{article}
\usepackage{ijcai19}
\usepackage{booktabs}
\usepackage{latexsym}
\usepackage{graphicx}
\usepackage{algorithm}
\usepackage[noend]{algorithmic}
\usepackage{amsmath}
\usepackage{amssymb}
\usepackage{url}
\usepackage{color,ifpdf}

\newcommand{\be}{\begin{equation}}
\newcommand{\ee}{\end{equation}}

\usepackage{times}
\usepackage{soul}
\usepackage{url}
\usepackage[hidelinks]{hyperref}
\usepackage[utf8]{inputenc}
\usepackage[small]{caption}
\usepackage{graphicx}
\usepackage{amsmath}
\usepackage{booktabs}
\usepackage{algorithm}
\usepackage{algorithmic}
\def\RECEIVED{\REQUIRE}

\def\PROCEDURE{\ENSURE}

\newcommand{\cA}{{\mathcal A}}
\newcommand{\cX}{{\mathcal X}}
\newcommand{\cR}{{\mathcal R}}
\newcommand{\cD}{{\mathcal D}}
\newcommand{\cmax}{q}
\newcommand{\cinf}{q_\infty}

\newcommand{\bu}{\mbox{\bf u}}
\newcommand{\bw}{\mbox{\bf w}}
\newcommand{\bz}{\mbox{\bf z}}

\newcommand{\ShareCPA}{\mbox{sCPA}}
\newcommand{\ShareUpperBound}{\mbox{sUB}}

\newcommand{\ZZ}{\mathbb{Z}}
\newcommand{\cE}{{\mathcal E}}
\newcommand{\comment}[1]{}

\newtheorem{theorem}{Theorem}[section]
\newtheorem{lemma}[theorem]{Lemma}

\title{A Privacy Preserving Collusion Secure DCOP Algorithm}

\author{
Tamir Tassa$^1$
\and
Tal Grinshpoun$^2$\and
Avishay Yanai$^3$
\affiliations
$^1$The Open University, Ra'anana, Israel\\
$^2$Ariel University, Ariel, Israel\\
$^3$Bar-Ilan University, Ramat Gan, Israel
\emails
tamirta@openu.ac.il, talgr@ariel.ac.il, ay.yanay@gmail.com
}

\begin{document}

\maketitle

\begin{abstract}
In recent years, several studies proposed privacy-preserving algorithms for solving Distributed Constraint Optimization Problems (DCOPs).
All of those studies assumed that agents do not collude.
In this study we propose the first privacy-preserving DCOP algorithm
that is immune to coalitions, under the assumption of honest majority.
Our algorithm -- PC-SyncBB -- is based on the classical Branch and Bound DCOP algorithm. It offers constraint, topology and decision privacy. We evaluate its performance on different benchmarks, problem sizes, and constraint densities. We show that achieving security against coalitions is feasible. As all existing privacy-preserving DCOP algorithms base their security on assuming solitary conduct of the agents, we view this study as an essential first step towards lifting this potentially harmful assumption in all those algorithms.
\end{abstract}

\section{Introduction}
Constraint optimization~\cite{MesLar95} is a powerful framework for describing optimization problems in terms of constraints. In many practical domains, such as Meeting Scheduling~\cite{MaTBPV04}, Mobile Sensor nets~\cite{FarinelliRPJ08}, and
the Internet of Things \cite{Lezama2017}, the constraints are enforced by distinct participants (agents). Hirayama and Yokoo~\shortcite{hirayama97distributed} termed such problems as Distributed Constraint Optimization Problems (DCOPs). Various algorithms for solving DCOPs have been proposed, some of which are complete~\cite{GerMeiZiv09,hirayama97distributed,MailLesser04,ModiSTY05,PetcuF05,YeohFK10},
and some are incomplete~\cite{FarinelliRPJ08,KatagishiP07,ottens2017duct,ZhangWXW05}.

The main motivation for DCOP research stems from the inherent distributed structure of many real-world problems, and the \emph{privacy} concerns that are associated with this distribution.
L{\'e}aut{\'e} and Faltings \shortcite{LeaFal13} offered the basic definitions of privacy in this framework. The four notions of privacy that they describe are: agent privacy, topology privacy, constraint privacy and decision privacy (see Section \ref{pre}).
Several studies considered a solution of DCOPs in a manner that preserves (some of) those privacy types.

This line of research began with the work of Silaghi and Mitra~\shortcite{Silaghi04}. They proposed a privacy-preserving solution to
Distributed Weighted Constraint Satisfaction Problems (DisWCSPs); those are distributed problems that are similar to DCOPs, but differ from them in the distribution model and, consequently, in the related privacy targets. Their solution is strictly limited to small scale problems since it depends on an exhaustive search over all possible assignments. As their solution is based on the BGW protocol \cite{BGW88}, it is immune against coalitions that involve less than half of the agents.

All of the subsequent studies considered DCOPs. 
The main motif in those studies was to develop privacy-preserving versions of existing algorithms. Greenstadt et al.~\shortcite{GreGroSmi07} devised a version of the DPOP algorithm \cite{PetcuF05}, called SSDPOP.
L{\'e}aut{\'e} and Faltings~\shortcite{LeaFal13} proposed three privacy-preserving versions of DPOP
that differ in their privacy guarantees and in their runtime performance.
Grinshpoun and Tassa~\shortcite{GT16} developed P-SyncBB, a privacy-preserving version of the complete search algorithm SyncBB \cite{hirayama97distributed}.
Tassa et al.~\shortcite{TassaGZ17} presented P-Max-Sum, a privacy-preserving version of the incomplete inference-based Max-Sum algorithm \cite{FarinelliRPJ08}.
Lastly, Grinshpoun et al.~\shortcite{grinshpoun2019privacy} devised P-RODA, a secure implementation of region-optimal algorithms.

All of the above described works based their security on assuming {\em solitary} conduct of the agents.
Alas, subsets of agents may try to collude and combine the information which they have
in order to infer information on other agents.
In this paper we suggest the first privacy-preserving DCOP algorithm that is immune against such coalitions. We depart from the SyncBB algorithm and devise a privacy-preserving algorithm that simulates its operation and provides topology, constraint and decision privacies, even in the presence of a coalition of agents, under the assumption of an honest majority (i.e., the size of the coalition is smaller than half the number of agents).


\section{Definitions and assumptions}\label{pre}
A Distributed Constraint Optimization Problem (DCOP)
is a tuple $\langle \cA,\cX,\cD,\cR \rangle$ where
$\cA$ is a set of agents $A_1,A_2, \ldots,A_n$, $\cX$ is a
set of variables $X_1,X_2,\ldots,X_m$, $\cD$ is a set of finite domains $D_1, D_2, \ldots,D_m$, and $\cR$ is a set of relations (constraints).
Each variable $X_i$ takes values in the domain $D_i$, and it is held by a single agent.
Each constraint $C \in \cR$ defines a non-negative cost for every possible value combination of a set of variables, and is of the form
$C : D_{i_1} \times  \cdots \times  D_{i_k} \rightarrow [0,\cmax]$, for some $1 \leq i_1 < \cdots < i_k \leq m$, and a publicly known maximal constraint cost $q$.

An {\em assignment} is a pair including a variable, and a value from that variable's domain. We denote by $a_i$ the value assigned to the
variable $X_i$. A {\em partial assignment} (PA) is a set of assignments in which each variable appears at most once. A constraint $C \in \cR $ is {\em applicable} to a PA if all variables that are constrained by $C$ are included in the PA.
The cost of a PA is the sum of all applicable constraints to the PA. A {\em full assignment} is a partial assignment that includes all of the variables.
The goal in Constraint Optimization Problems is to find a full assignment of minimal cost.

For simplicity, we assume that each agent holds exactly one variable, i.e., $n=m$. We let $n$ denote hereinafter the number of agents and the number of variables. We consider a binary version of DCOPs, in which every $C\in \cR$ constraints exactly two variables and takes the form
$C_{i,j} : D_i \times D_j  \rightarrow [0,q]$.
These assumptions are customary in DCOP literature \cite{ModiSTY05,PetcuF05}.

L{\'e}aut{\'e} and Faltings \shortcite{LeaFal13} have distinguished between four notions of privacy. The notions of privacy which our proposed algorithm respects are:
(a) \emph{Topology privacy} -- hiding from each agent the topological structures in the constraint graph beyond his\footnote{We use the masculine form for simplicity.} own direct neighborhood in the graph; (b) \emph{Constraint privacy} -- hiding from each agent the constraints in which he is not involved; and (c)
\emph{Decision privacy} -- hiding from each agent the final assignments to other variables.
\comment{
L{\'e}aut{\'e} and Faltings \shortcite{LeaFal13} have distinguished between four notions of privacy. The notions of privacy which our proposed algorithm respects are the following:
\begin{itemize}
	\item \emph{Topology privacy} -- hiding from each agent the topological structures in the constraint graph beyond his own direct neighborhood in the graph.
  \item \emph{Constraint privacy} -- hiding from each agent the constraints in which he is not involved. Namely, agent $A_k$ should not know anything about $C_{i,j}(\cdot,\cdot)$ if $k \notin \{i,j\}$.
	\item \emph{Decision privacy} -- hiding from each agent the final assignments to other variables.
\end{itemize}
}

Like in all prior art on privacy-preserving DCOP algorithms, we too assume that the agents are semi-honest, namely,
they follow the prescribed protocol but try to glean more
information than allowed from the protocol transcript. In contrast to prior art, we assume that (less than half of the) agents may collude in order to combine their inputs and messages received during the execution of the protocol, for the purpose of extracting private information on other agents.

\section{A Secure Synchronous Branch and Bound}\label{main}
Synchronous Branch-and-Bound (SyncBB) \cite{hirayama97distributed} was the first complete algorithm for solving DCOPs. SyncBB operates in a completely sequential manner, a fact that inherently renders its synchronous behavior.
It assumes a static public ordering of the agents, $A_1,\ldots,A_n$.
The search space of the problem is traversed by each agent assigning a value to his variable and passing the \emph{current partial assignment} (CPA) to the next agent in the order, along with the current cost of the CPA. After an agent completes assigning all values in the domain to his variable, he \emph{backtracks}, i.e., he sends the CPA back to the preceding agent. To prevent exhaustive traversal of the entire search space, the agents maintain an \emph{upper bound}, which is the cost of the best solution that was found thus far. The algorithm keeps comparing the costs of CPAs and the current upper bound, in order to \emph{prune} the search space.

Herein we devise a secure implementation of SyncBB, called PC-SyncBB (Privacy-preserving and Collusion-resistant SyncBB).
Another secure implementation of SyncBB, called P-SyncBB, was previously introduced by Grinshpoun and Tassa~\shortcite{GT14,GT16}.
The two algorithms are fundamentally different.
\comment{
(1) PC-SyncBB is secure against coalitions; P-SyncBB's security is jeopardized already when two agents collude.

(2) As opposed to P-SyncBB, in PC-SyncBB the cost of CPAs and the upper bound are never revealed to anyone,
and the current assignment of any variable is known only to the agent that controls it.

(3) The cryptographic approach in PC-SyncBB differs significantly from the one underlying P-SyncBB.
}
While in P-SyncBB agents are exposed to sensitive information such as assignments of other agents, costs of CPAs, and the value of the upper bound, PC-SyncBB totally avoids such information disclosure. Hence, the outline of PC-SyncBB is simpler than that of P-SyncBB, because there is no need to implement mechanisms for preventing illegal inferences that can be deduced from such information.
On the other hand, as in PC-SyncBB much less information is revealed, and as PC-SyncBB is designed to be resistant to coalitions
(while P-SyncBB's security is jeopardized already when two agents collude),
the secure multi-party computational tasks in PC-SyncBB are much harder. Hence, the cryptographic approach taken in PC-SyncBB is completely different,
and it is much more involved than the corresponding one in P-SyncBB.

\subsection{Preliminaries}\label{sec1pre}

\noindent
{\bf General assumptions and notations.}
The design of PC-SyncBB is based on the following general assumptions:

(1) There is a static public ordering of
the agents, $A_1,\ldots,A_n$.

(2) The upper bound on the cost of any possible solution is
$\cinf:={n \choose 2}\cmax +1$, and it is known to all agents. In addition, all agents agree upfront on an integer $S$ greater than $2 \cinf$.

(3) For every pair of indices $1 \leq t < k \leq n$, $\Gamma(t,k)$ is a Boolean predicate that equals {\bf true} iff
$X_t$ and $X_k$ are constrained. Then, $I_k^-:=\{ t: 1 \leq t < k \mbox{ and } \Gamma(t,k) \}$ and
$I_k^+:=\{ t: k < t \leq n \mbox{ and } \Gamma(k,t) \}$ are sets containing the indices of all agents that precede/follow $A_k$ in the order and whose variable is constrained with $X_k$. We also let $I_k:=I_k^- \cup I_k^+$.

\smallskip
\noindent
{\bf Value ordering.}
Each agent $A_k$ maintains two value orderings over his domain $D_k$.
Each of those orderings can be described by a vector of length $|D_k|$.
The first ordering, denoted $\bu_k$, is fixed and known to all agents $A_t$ such that $t \in I_k$.
Then if $A_t$ and $A_k$ are constrained, they can describe their constraint $C_{t,k}$ as a matrix $M_{t,k}$ of $|D_t|$ rows and $|D_k|$ columns, where the value in the $r$-th row and $s$-th column is
\be M_{t,k}(r,s)=C_{t,k}(\bu_t(r),\bu_k(s))\,.
\label{mdef}\ee
The second ordering, denoted $\bw_k$, is generated at random by $A_k$ whenever he begins a new traversal over his domain.
That ordering, which is kept secret from all other agents, determines the order in which that agent will scan the values in his domain during that stage of the search. Agent $A_k$ generates such an ordering each time a CPA is passed to him from the preceding agent $A_{k-1}$.

\smallskip
\noindent
{\bf Internal variables.}
Every agent $A_k$ maintains the following variables:

(1) $\ShareCPA_k$ is an array of length $n$ that holds additive shares in the cost of the CPA.
Assume that agents $A_t$ and $A_k$ are constrained and that $C_{t,k}$ is applicable to the CPA. Then the cost of the CPA includes,
as one of its addends, the value $C_{t,k}(X_t,X_k)$. In such a case $\ShareCPA_k(t)$ and $\ShareCPA_t(k)$ will both store random values
in $\ZZ_S$ so that
\be \ShareCPA_t(k) +\ShareCPA_k(t) = C_{t,k}(X_t,X_k) \mbox{ mod } S \,. \label{costkt}\ee
If, on the other hand, $C_{t,k}$ is {\em not} applicable to the CPA (i.e. the CPA does not include $X_k$ or $X_t$ or both), then $\ShareCPA_k(t) =\ShareCPA_t(k) = 0$.
In view of the above, the overall cost of the CPA, at any stage of the algorithm's run, equals
\be Cost(CPA) = \sum_{k=1}^n \sum_{t \in I_k} \ShareCPA_k(t) \mbox{ mod } S \,. \label{costcpa}\ee

(2) $\ShareUpperBound_k $ holds
an additive share in the upper bound (the cost of the best full assignment that was discovered thus far).
Each such share is random and uniformly distributed over $\ZZ_S$.
At any stage of the algorithm's run,
\be UpperBound = \sum_{k=1}^n \ShareUpperBound_k \mbox{ mod } S \,. \label{upperbound}\ee

(3) $p_k$ is a pointer to a value in the ordering $\bw_k$. The current assignment to $X_k$ is given by $\bw_k(p_k)$.

(4) $OptimalSetting_k$ stores the assignment to $X_k$ in the currently best full assignment that was found thus far.

\subsection{The PC-SyncBB algorithm}\label{sec1alg}
\medskip\noindent
The PC-SyncBB algorithm is given in Algorithm \ref{alg:PSBB}, which we proceed to describe.

\begin{algorithm}[h!]
\caption{ \textbf{--} PC-SyncBB (executed by agent $A_k$)}
\label{alg:PSBB}
\small

\begin{algorithmic}[1]
\PROCEDURE \textbf{init}
\STATE $\ShareCPA_k(t) \leftarrow 0$ for all $1 \leq t \leq n$
\STATE $p_k \leftarrow 0$
\IF {$k>1$}
    \STATE $\ShareUpperBound_k \leftarrow 0$
\ELSE
	\STATE $\ShareUpperBound_k \leftarrow \cinf$
	\STATE assign\_CPA()
\ENDIF
\item[]

\PROCEDURE \textbf{assign\_CPA}
\IF {$p_k =0$}
\STATE Generate a new random ordering of $D_k$ into $\bw_k$
\ENDIF
\STATE $p_k \leftarrow p_k +1 $
\IF {$p_k > |D_k|$}
\STATE backtrack()
\ELSE
\STATE $X_k \leftarrow v:=\bw_k(p_k)$
\STATE update\_shares\_in\_CPA$(k,v)$
\IF {$k=n$}
	\IF {compare\_CPA\_cost\_to\_upper\_bound() $=$ {\bf true}}
	\STATE broadcast({\bf NEW\_OPTIMUM\_FOUND})
    \ENDIF
    \STATE assign\_CPA()
\ELSE
	\IF {compare\_CPA\_cost\_to\_upper\_bound() $=$ {\bf false}}
		\STATE assign\_CPA()
	\ELSE
		\STATE send({\bf CPA\_MSG}) to $A_{k+1}$
	\ENDIF
\ENDIF
\ENDIF
\item[]

\PROCEDURE \textbf{backtrack}
\IF {$k>1$}
	\STATE $\ShareCPA_k(t) \leftarrow 0$ for all $t \in I_k^-$
    \STATE send({\bf ZERO\_SHARE\_MSG},$k$) to $A_t$ for all $t \in I_k^-$
	\STATE send({\bf BACKTRACK\_MSG}) to $A_{k-1}$
\ELSE
     \STATE broadcast({\bf COMPLETE})
\ENDIF
\item[]

\RECEIVED ({\bf NEW\_OPTIMUM\_FOUND}) \textbf{do}
\STATE $\ShareUpperBound_k \leftarrow \sum_{t \in I_k} \ShareCPA_k(t)$
\STATE $OptimalSetting_k \leftarrow X_k$

\RECEIVED ({\bf CPA\_MSG}) \textbf{do}
\STATE $p_k \leftarrow 0$
\STATE assign\_CPA()

\RECEIVED ({\bf ZERO\_SHARE\_MSG},$k'$) \textbf{do}
\STATE $\ShareCPA_k(k') \leftarrow 0$

\RECEIVED ({\bf BACKTRACK\_MSG}) \textbf{do}
\STATE assign\_CPA()

\RECEIVED ({\bf COMPLETE}) \textbf{do}
\STATE $X_k \leftarrow OptimalSetting_k$
\STATE Terminate

\end{algorithmic}

\end{algorithm}

\medskip\noindent
{\bf The procedure init.} Every agent $A_k$ initializes all entries in his vector $\ShareCPA_k$ as well as $p_k$ to zero (Lines 1-2).
Then, every agent $A_k$, $k>1$, initializes $\ShareUpperBound_k$ to zero, while $A_1$ initializes it to $\cinf$ (Lines 3-6). Such settings imply that  $\sum_{k=1}^n \ShareUpperBound_k = \cinf$ mod $S$,
in agreement with Eq. (\ref{upperbound}) (since the initial upper bound is set to $\cinf$).
Finally, the procedure init triggers the search by having $A_1$ call the procedure assign\_CPA (Line 7).

\medskip\noindent
{\bf The procedure assign\_CPA.}
If this procedure is called when $p_k=0$, it means that $A_k$ now begins a new traversal over his domain. Hence, in such a case he generates a new random ordering, $\bw_k$, of $D_k$ (Lines 8-9).
In order to move to the next value in $\bw_k$, $A_k$ increments the pointer $p_k$ (Line 10).
If $p_k$ becomes greater than $|D_k|$ it means that the domain $D_k$ was already fully scanned, so $A_k$ performs the procedure backtrack (discussed below) in order to return the search torch back to the preceding agent $A_{k-1}$ (Lines 11-12).
Otherwise, $A_k$ assigns $v:=\bw_k(p_k)$ to $X_k$ (Line 14).
Consequently, as $X_k$ has a new value, the CPA's cost is changed, so new random shares of that cost must be computed.
This is done by calling the sub-protocol update\_shares\_in\_CPA$(k,v)$ (Line 15),
which recomputes $\ShareCPA_k(t)$ and $\ShareCPA_t(k)$, for all $t \in I_k^-$, so that the right-hand side of Eq. (\ref{costcpa}) equals the new CPA's cost. (We discuss that sub-protocol in Section~\ref{sec-smc1}.)

We now separate the discussion according to the index $k$ of the operating agent.
If $k=n$, then a new full assignment is reached. It is needed to compare its cost, which equals
$\sum_{k=1}^n \sum_{t \in I_k} \ShareCPA_k(t)$ mod $S$, Eq. (\ref{costcpa}), to the current upper bound, $\sum_{k=1}^n \ShareUpperBound_k$ mod $S$, (Eq. (\ref{upperbound})).
This comparison must be done in a secure manner. To that end, $A_n$ invokes compare\_CPA\_cost\_to\_upper\_bound (Line 17),
a secure multi-party sub-protocol
that we discuss in Section~\ref{sec-smc2}. It returns {\bf true} if the cost of the current full assignment is lower than the upper bound, namely, if
\be \sum_{k=1}^n \sum_{t \in I_k} \ShareCPA_k(t) \mbox{ mod } S < \sum_{k=1}^n \ShareUpperBound_k \mbox{ mod } S\,,\label{mainineq}\ee
and {\bf false} otherwise.
If the current full assignment does improve the upper bound, then
$A_n$ broadcasts the message {\bf NEW\_OPTIMUM\_FOUND} (Line 18). Upon receiving such a message, every agent $A_k$ stores
the sum of his current shares,
$\sum_{t \in I_k} \ShareCPA_k(t)$,
in $\ShareUpperBound_k$ and he also stores the current assignment of $X_k$ in $OptimalSetting_k$ (Lines 31-32).
Finally, whether the current full assignment is a new optimum or not, $A_n$ calls the procedure assign\_CPA again in order to test the next value in his domain (Line 19).

If $k<n$, the agents examine the possibility to prune the search space: they first check whether the CPA's cost is already greater than or equal to the upper bound,
by invoking compare\_CPA\_cost\_to\_upper\_bound (Line 21).
If it returns {\bf false} then Eq. (\ref{mainineq}) does not hold, i.e., the cost of the CPA is already greater than or equal to the upper bound. In such a case there is no point in pursuing the current path in the search space, so $A_k$
calls the procedure assign\_CPA again in order to test the next value in his domain (Line 22). Otherwise, $A_k$ passes the torch onward to $A_{k+1}$ (by sending him the message {\bf CPA\_MSG} in Line 24) in order to continue the search over CPAs with the current $k$-prefix. When $A_{k+1}$ receives the message
{\bf CPA\_MSG}, he zeroes the pointer $p_{k+1}$ to his domain $D_{k+1}$, in order to start traversing all values in $D_{k+1}$ as possible extensions to the current $k$-CPA, and then he calls the procedure assign\_CPA (Lines 33-34).

\medskip\noindent
{\bf The procedure backtrack.}
When agent $A_k$, $k>1$, executes the procedure backtrack, he does two things. First, he zeros all entries in $\ShareCPA_k$ (Line 26) and sends a {\bf ZERO\_SHARE\_MSG} message, with his index $k$, to all agents that precede him and are constrained with him (Line 27). Any such agent, upon receiving the {\bf ZERO\_SHARE\_\\MSG} message, zeroes the relevant share in his own array (Line 35). As a result of the above two actions, Eq. (\ref{costcpa}) still holds for the reduced CPA that is obtained after this backtracking.
Afterwards, $A_k$ sends a {\bf BACKTRACK\_MSG} message to $A_{k-1}$ (Line 28). When the latter receives that message, he calls assign\_CPA in order to change the assignment of his variable to the next value in his domain and proceed the search with the new modified CPA (Line 36).

When $A_1$ performs backtrack, it means that he completed a traversal of $D_1$, and, consequently, the entire search space ($D_1 \times \cdots \times D_n$) was scanned. Therefore, the algorithm terminates with the last optimum found being the global optimum. In such a case $A_1$ broadcasts the message {\bf COMPLETE} (Line 30). When receiving such a message, every agent
$A_k$ assigns to his variable $X_k$ the value $OptimalSetting_k$ (which was his assignment in the last optimal solution that was found) and then he terminates (Lines 37-38).

\subsection{The sub-protocol update\_shares\_in\_CPA}\label{sec-smc1}
Before starting PC-SyncBB, each of the agents $A_k$, $k<n$, creates a key pair in a Paillier cipher \cite{pai} and sends the corresponding public key to $A_t$ for all $t \in I_k^+$.
Denote by $\cE_k$ the encryption function in $A_k$'s cipher and by $\nu_k$ the corresponding modulus. Then $\cE_k$ is a function from
$\ZZ_\nu$ to $\ZZ^*_{\nu^2}$ and it is additively homomorphic, in the sense that for every two plaintexts $x$ and $y$, $\cE_k(x+y)=\cE_k(x) \cdot \cE_k(y)$, where addition is modulo $\nu$ and multiplication is modulo $\nu^2$.
The Paillier cipher is probabilistic, in the sense that the encryption function depends also on a random string (so that every plaintext $x$ has a large number of possible ciphertexts $\cE_k(x)$). It is known to be semantically secure \cite{GoldwasserM82}.

After creating $\cE_k$, $A_k$ computes a vector $\bz_k^1$ of length $|D_k|$ where $\bz_k^1(1)=\cE_k(1)$ and $\bz^1_k(i)=\cE_k(0)$ for all $2 \leq i\leq |D_k|$. It is important to compute the latter $|D_k|-1$ encryptions with $|D_k|-1$ independently selected random strings.
Then, $A_k$ defines the vectors
$ \bz_k^i = CRS \left( \bz^{i-1}_k \right)$, for $2 \leq i \leq |D_k|$, where $CRS(\cdot)$ is a circular right-shift by one position of the vector entries. Hence, $\bz_k^i$ encrypts the vector $(0,\ldots,0,1,0,\ldots,0)$ where the 1 appears in the $i$th entry,
$1 \leq i\leq |D_k|$. Given the manner in which those vectors were computed and the probabilistic and semantic security properties of the Paillier cipher,
a polynomially-bounded adversary who
gets any random sequence of those vectors (i.e. $\bz_k^{i_1}, \bz_k^{i_2},\ldots$) will not be able to distinguish between the $\cE_k(1)$ and the $\cE_k(0)$ entries in them (with a non-negligible probability of success).

We are now ready to describe the sub-protocol update\_shares\_in\_CPA (Algorithm \ref{alg:updateshares}).
It is triggered by $A_k$ whenever he assigns a new value $v$ to his variable, $X_k$. When that happens, it is needed to update the shares of all agents $A_1,\ldots,A_k$ so that the validity of Eq. (\ref{costcpa}) is maintained.
The shares that should be modified in wake of such an assignment are $\ShareCPA_k(t)$ and $\ShareCPA_t(k)$
for all $t \in I_k^-$. Those shares will be modified so that, in view of Eq. (\ref{costkt}),
the sum of $\ShareCPA_k(t)$ and $\ShareCPA_t(k)$, for any fixed $t \in I_k^-$,
will equal $C_{t,k}(X_t,X_k)$ for the current assignments of $X_t$ and $X_k$ ($X_k$'s assignment equals $v$, and it is passed to the sub-protocol as an input).

Assume that $t \in I_k^-$. Then the contribution of the pair $X_t$ and $X_k$ to the CPA is $M_{t,k}(r,s)$,
where $\bu_t(r)=X_t$ and $\bu_k(s)=X_k$
(see Eq. (\ref{mdef})).
Recall that $A_t$ does not know $s$ while $A_k$ does not know $r$. In order to compute the new respective shares, $\ShareCPA_k(t)$ and $\ShareCPA_t(k)$, so that Eq. (\ref{costkt}) holds, these two agents perform the following computation.

When $A_t$ performed last time the procedure assign\_CPA and set there the current assignment to $X_t$, he called
update\_shares\_in\_CPA (Algorithm \ref{alg:updateshares}), see Line 15 in PC-SyncBB. In Line 8 of Algorithm \ref{alg:updateshares} he
sent to all agents in $I_t^+$ the vector $\bz_t^j$ which encodes his current assignment.
Going back to the present, when $A_k$ executes update\_shares\_in\_CPA
he holds a vector $\bz_t$ that he received from $A_t$ for every $t \in I_k^-$.
That vector equals $\bz_t^r$, where
$r$ is the index in $\bu_t$ in which the current assignment to $X_t$ is stored.
Even though $A_k$ cannot infer from $\bz_t$ the current value of $X_t$,
he can still correctly update his shares vis-a-vis $A_t$. To that end,
he computes
\be y_t:=\prod_{i=1}^{|D_t|} \bz_t(i)^{[(M_{t,k}(i,s) - \rho) \mbox{ mod } S]} \,, \label{yt}\ee
where $s$ is the index of the entry in $\bu_k$ that holds $v$ -- the current assignment to $X_k$, and
$\rho$ is a value selected uniformly at random (independently for each $A_t$) from $\ZZ_S$.
The key observation here is the following.

\begin{lemma}\label{lm1}
The homomorphism of $\cE_t$ implies that $y_t= \cE_t\left( [(M_{t,k}(r,s) - \rho) \mbox{ mod } S] \right) $.
\end{lemma}

Next, $A_k$ sends $y_t$ to $A_t$ who decrypts it and stores it in $\ShareCPA_t(k)$. In view of Lemma \ref{lm1}, $A_t$ obtains $\ShareCPA_t(k) = (M_{t,k}(r,s) - \rho) \mod S$ whereas $A_k$ sets $\ShareCPA_k(t)=\rho \mod S$ (Algorithm \ref{alg:updateshares}, Lines 5-6).
Those two uniformly random shares satisfy
$\ShareCPA_t(k) + \ShareCPA_k(t) = M_{t,k}(r,s) \mod S$, which fulfils
the required equality in Eq. (\ref{costkt}).

The above described updates are carried out by $A_k$ and $A_t$ for all $t \in I_k^-$. After completing all those updates, the updated shares
satisfy Eq. (\ref{costcpa}).

\begin{algorithm}[h!!!]
\caption{ The sub-protocol update\_shares\_in\_CPA}

\label{alg:updateshares}
\small

\begin{algorithmic}[1]
\REQUIRE $k$, the index of the agent $A_k$ that invokes the procedure, and $v$, $A_k$'s current assignment
\FORALL {$t \in I_k^-$}
\STATE $A_k$ selects uniformly at random $\rho \in \ZZ_S$.
\STATE $A_k$ computes $y_t$ as given in Eq. (\ref{yt}), where $\bz_t$ is the vector that $A_k$ received from $A_t$ in the last time.
\STATE $A_k$ sends the computed $y_t$ to $A_t$.
\STATE $A_t$ sets $\ShareCPA_t(k) \leftarrow \cE_t^{-1}(y_t)$.
\STATE $A_k$ sets $\ShareCPA_k(t) \leftarrow \rho$.
\ENDFOR
\IF {$k<n$}
\STATE $A_k$ sends to all $A_t$ where $t \in I_k^+$ the vector $\bz_k^j$ where $j$ is the index for which
$\bu_k(j)=v$.
\ENDIF
\end{algorithmic}

\end{algorithm}

\subsection{compare\_CPA\_cost\_to\_upper\_bound}\label{sec-smc2}
The sub-protocol compare\_CPA\_cost\_to\_upper\_bound verifies the inequality in Eq. (\ref{mainineq}).
Agent $A_k$, $1 \leq k \leq n$, holds two integers modulo $S$: $ a_k:=\sum_{t \in I_k} \ShareCPA_k(t) \mbox{ mod } S $
and $b_k:=\ShareUpperBound_k \mbox{ mod } S$.
The goal is to determine whether $\alpha:=\sum_{k=1}^n a_k \mbox{ mod } S$ is smaller than
$\beta:=\sum_{k=1}^n b_k \mbox{ mod } S$ or not.

To this end, we make use of a secure multi-party computation (MPC) protocol \cite{yao,BMR}. An MPC protocol for some function $f$ allows a set of $n$ distrustful parties $A_1,\ldots,A_n$, where $A_i$ possesses a private input $x_i$, to compute $y\gets f(x_1,\ldots,x_n)$,
while preserving the privacy of the parties. Namely, at the end of the protocol the parties learn $y$, but nothing beyond that on inputs of other parties. Almost all practical MPC protocols work with the underlying Boolean/Arithmetic circuit as the computational model. Therefore, to securely compute the function $f$, the parties first need to agree on a Boolean/Arithmetic circuit $C$ that implements $f$. The runtime of such computations depends on $n$ and the size $|C|$ of the circuit $C$ (the number of gates in it).

In PC-SyncBB, we managed to shrink the usage of a general purpose MPC protocol to the computation of Eq. (\ref{mainineq}). Specifically, we use the protocol of Ben-Efraim and Omri \shortcite{BO17} for the function $f(x_1,\ldots,x_n)$ where $A_k$'s input is $x_k=b_k-a_k \mod S$ and the function $f$ returns {\bf true} if
Eq. (\ref{mainineq}) holds and {\bf false} otherwise.

The protocol of Ben-Efraim and Omri \shortcite{BO17} is secure under the assumption that less than $n/2$ parties collude.
It proceeds in two phases, called ``offline'' and ``online''. In the offline phase the parties do not yet know their inputs, but
can prepare the raw materials required for the computation. The online phase begins once the parties know their inputs. In practice, this allows the parties to perform the offline phase in advance, even without having their inputs (or even knowing the domains), and perform the fast online phase once the inputs are ready. Due to page limitations, we defer the detailed description of this sub-protocol to the full version.

%
%


\newpage
\subsection{Properties of PC-SyncBB}\label{properties}
The main properties of this algorithm are stated below.

\begin{theorem}
PC-SyncBB is complete and sound.
\end{theorem}

{\em Proof.}
The completeness of PC-SyncBB follows from the exhaustive search structure. Only partial assignments whose cost reach the upper bound are not extended
and therefore it is
guaranteed that the algorithm finds an optimal solution.
Termination also follows from the exhaustive structure of the Branch-and-Bound algorithm in which no partial assignment can be explored twice.

PC-SyncBB is sound, in the sense that it outputs a correct solution, as implied by the
correctness of update\_shares\_in\_\\CPA (which guarantees that Eqs. (\ref{costcpa}) and (\ref{upperbound}) are always correct) and
compare\_CPA\_cost\_to\_upper\_bound (which guarantees the correctness of validating Eq. (\ref{mainineq})).
$\Box$

\begin{theorem}
PC-SyncBB provides constraint-, topology-, and assignment/decision-privacy. Even if any subset $\mathcal{B}  \varsubsetneq \mathcal{A}$ of agents collude, where $|\mathcal{B}|<n/2$, they would not be able to infer information on (values or existence of) constraints between two agents outside the coalition, or on value assignments or final decisions of such agents.
\end{theorem}

{\em Proof.} The only way in which privacy can be breached is through the data which is transmitted between agents.
In the main body of PC-SyncBB (Algorithm \ref{alg:PSBB}) the only data which the agents transmit between themselves are command messages.
Those messages convey information only with regard to the sizes of the variable domains, $|D_k|$, $1 \leq k \leq n$, but those domains are assumed to be publicly known anyway. Since the order in which each agent traverses his domain during the search is random and kept secret from all other agents, such messages do not include any information regarding the assignments, the final decisions, the constraints, or the constraint graph topology.

In addition to those command messages, information is exchanged also in the two sub-protocols.
In update\_shares\_in\_\\CPA, the agent $A_k$ receives from every $A_t$, where $t \in I_k^-$, his vector $\bz_t$. That vector is computed by $A_t$ whenever he assigns a new value from his domain to $X_t$. As each of those computations is made independently of previous computations, and as the Paillier cipher is semantically secure, $A_k$ cannot infer from $\bz_t$ any information on the current assignment of $A_t$. Moreover, as $A_t$ sends the same vector $\bz_t$ to all agents $A_k$, $k \in I_t^+$, upon their request, no coalition, of any size, can gain additional knowledge on $A_t$'s assignments.
Another place in update\_shares\_in\_CPA in which data is exchanged is in Line 4. There, agent $A_k$ sends to $A_t$ the value $y_t$, which includes the $\cE_t$-encryption of $ [(M_{t,k}(r,s) - \rho) \mbox{ mod } S] $. Since $\rho$ is selected uniformly at random from $\ZZ_S$, this value contains no information at all. Moreover, since $A_k$ selects in Line 3 an independent random $\rho$ for each $A_t$, also here there is no point in performing coalitions.

As for the compare\_CPA\_cost\_to\_upper\_bound sub-protocol, it is secure, under the assumption of honest majority, since it implements the Ben-Efraim-Omri protocol that was shown to be secure under that assumption \cite{BO17}.
$\Box$

It is important to note that, like {\em all} preceding papers on privacy-preserving solution of DCOPs, our algorithm does not guarantee {\em perfect} privacy, as it may leak some very benign information on the constraint graph topology. While achieving perfect privacy is possible, in theory, in any multi-party computation, it is very hard to do so while maintaining practicality. Hence, in almost all studies that deal with privacy-preserving solutions of practical problems, one accepts benign information leakages.

\section{Experimental evaluation}\label{exp}
We begin by evaluating the runtime of the compare\_CPA\_cost\_to\_upper\_bound sub-protocol, which is a central and computationally expensive part of PC-SyncBB.
For efficiency and reproducibility we used the original implementation of
Ben-Efraim and Omri
\shortcite{BO17}.\footnote{https://github.com/cryptobiu/Protocols/tree/master/Concrete\_\\Efficiency\_Improvements\_to\_Multiparty\_Garbling\_with\_an\_Honest\_\\Majority}
The executions were over LAN with EC2 machines of type {\sf c5.large} in Amazon's North Virginia data center with every agent running on a separate machine.
We measured performance for various values of $n$, where $\cmax$ (the maximum value of a single binary constraint) is set to 100.
Hence, as the maximum cost of any solution is $\cinf:={n \choose 2}\cmax +1$ and $S$ is set to be the smallest power of 2 greater than $2 \cinf$, then $n$ fully determines $\ell=\log S$ and, consequently, also the size of the circuit $C$ that the protocol uses.
Table \ref{tbl:mpc-results} gives, for each $n$, the bit-length $\ell$ of the agents' inputs, the overall circuit size (number of gates),
and the average runtimes over 100 executions for the offline and online phases of the protocol.
\comment{
The average running times over 100 executions for both the offline and online phases of the protocol, the bit-length of the agents' inputs and the overall circuit size 
are given in Table \ref{tbl:mpc-results}.\footnote{The original implementation uses only odd input sizes.}
}

\begin{table}[ht]
\small
\centering
\setlength\tabcolsep{4.2pt} 
\begin{tabular}{l|llllllll}
\hline
 $n$		 	& 5 	 & 7  		& 9  	& 11		&  13	& 15  	& 17  	 & 19 \\
 \hline
 $\ell$			& 11	 & 13 		& 13	& 14		& 14	& 15 	& 15	 & 16 \\
 $|C|$			& 184	 & 336 		& 448	& 610		& 732	& 924	& 1056	 & 1278 \\
 offline		& 6.7	 & 12.4		& 20.2	& 32.3		& 47.2	& 72.0	& 94.3	 & 135.3 \\
 online			& 0.51	 & 0.85		& 1.3	& 1.6		& 2.4	& 2.5	& 2.7	 & 3.6\\
 \hline	
\end{tabular}
\caption{Bit length, circuit size and runtime (msecs) of the Ben-Efraim-Omri as a function of $n$.}
\label{tbl:mpc-results}
\end{table}

Now we turn to the runtime performance evaluation of the full PC-SyncBB algorithm. In order to asses the toll of privacy preservation, we compare PC-SyncBB (offline+online) to other algorithms that maintain the Branch \& Bound structure -- P-SyncBB~\cite{GT16} that preserves privacy only under the assumption of non-colluding agents, and the basic insecure SyncBB~\cite{hirayama97distributed}. We also present separately the online requirements of PC-SyncBB.
The algorithms were implemented and executed in the AgentZero simulator \cite{lutati2014agentzero}, running on a hardware comprised of an Intel i7-6820HQ processor and 32GB memory, except for the calls to the compare\_CPA\_cost\_to\_upper\_bound procedure that were executed on the machines from Amazon's North Virginia data center, in order to simulate as realistically as possible a truly distributed environment.
We followed the {\em simulated time}~\cite{SultanikLR08} approach in all the subsequent experiments. The results are shown in a logarithmic scale and are the average over 50 problem instances (for each setting/benchmark).

\begin{figure}[h!!!]
\centering
\includegraphics[scale=0.25]{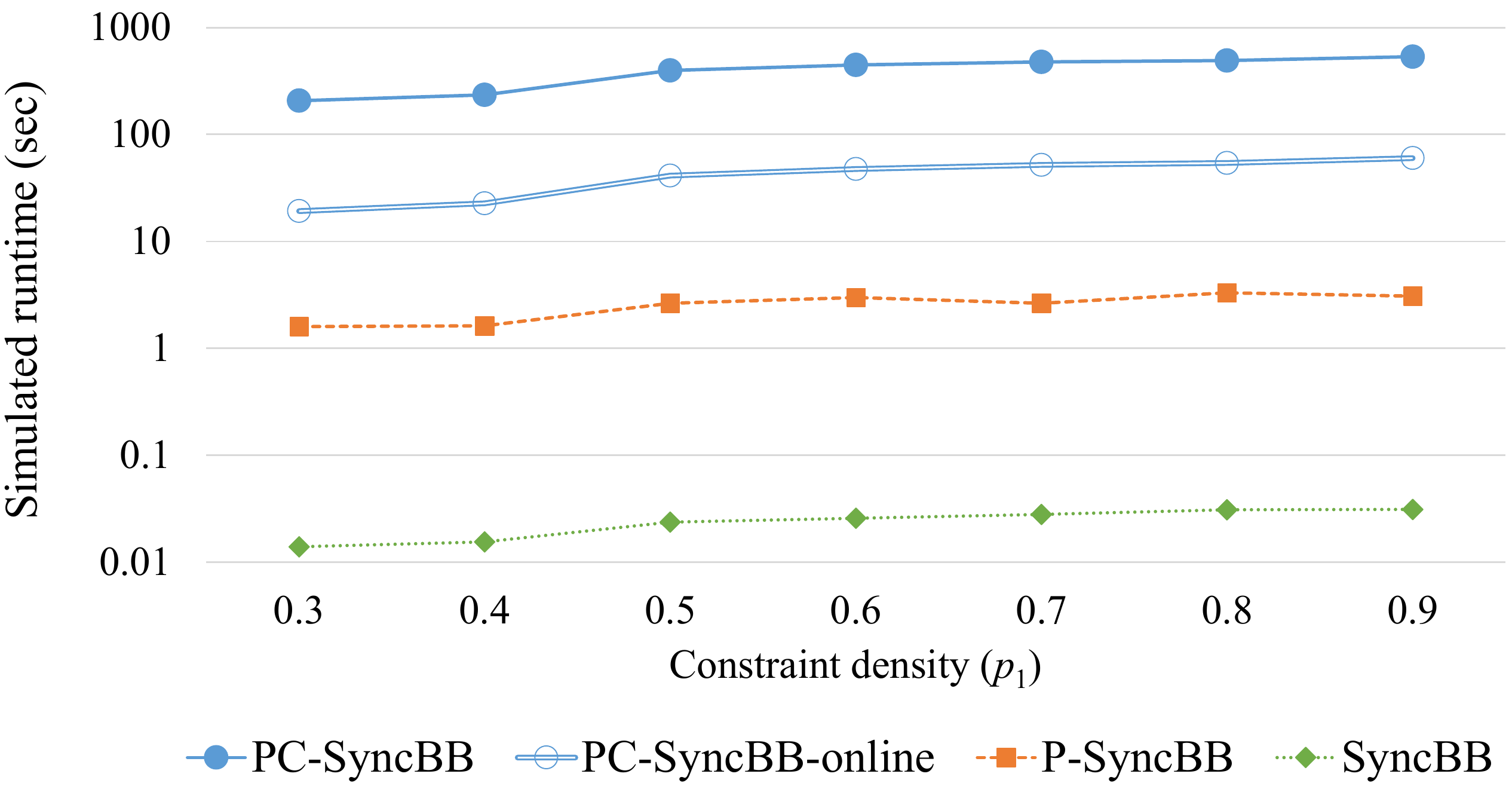}
\vspace{-6pt}
\caption{Varying $p_1$ (random DCOPs).}
\label{Fig:randomDensity}
\end{figure}

The first benchmark consists of unstructured randomly generated DCOPs on which we perform two experiments. In the first experiment, presented in Figure~\ref{Fig:randomDensity}, we fix the number of agents to $n=7$ and the domain sizes to 6, and vary the constraint density $0.3 \leq p_1 \leq 0.9$. (Using lower density values $p_1 < 0.3$ results in unconnected constraint graphs.)
It is clear that constraint density only mildly affects the runtime performance of all the evaluated algorithms. However, the toll of privacy preservation is evidently high, with each layer of protection adding about two orders of magnitude to the runtime. Specifically, the online part of PC-SyncBB requires about one order of magnitude more time than P-SyncBB.

\vspace{-2pt}

\begin{figure}[h!!!]
\centering
\includegraphics[scale=0.25]{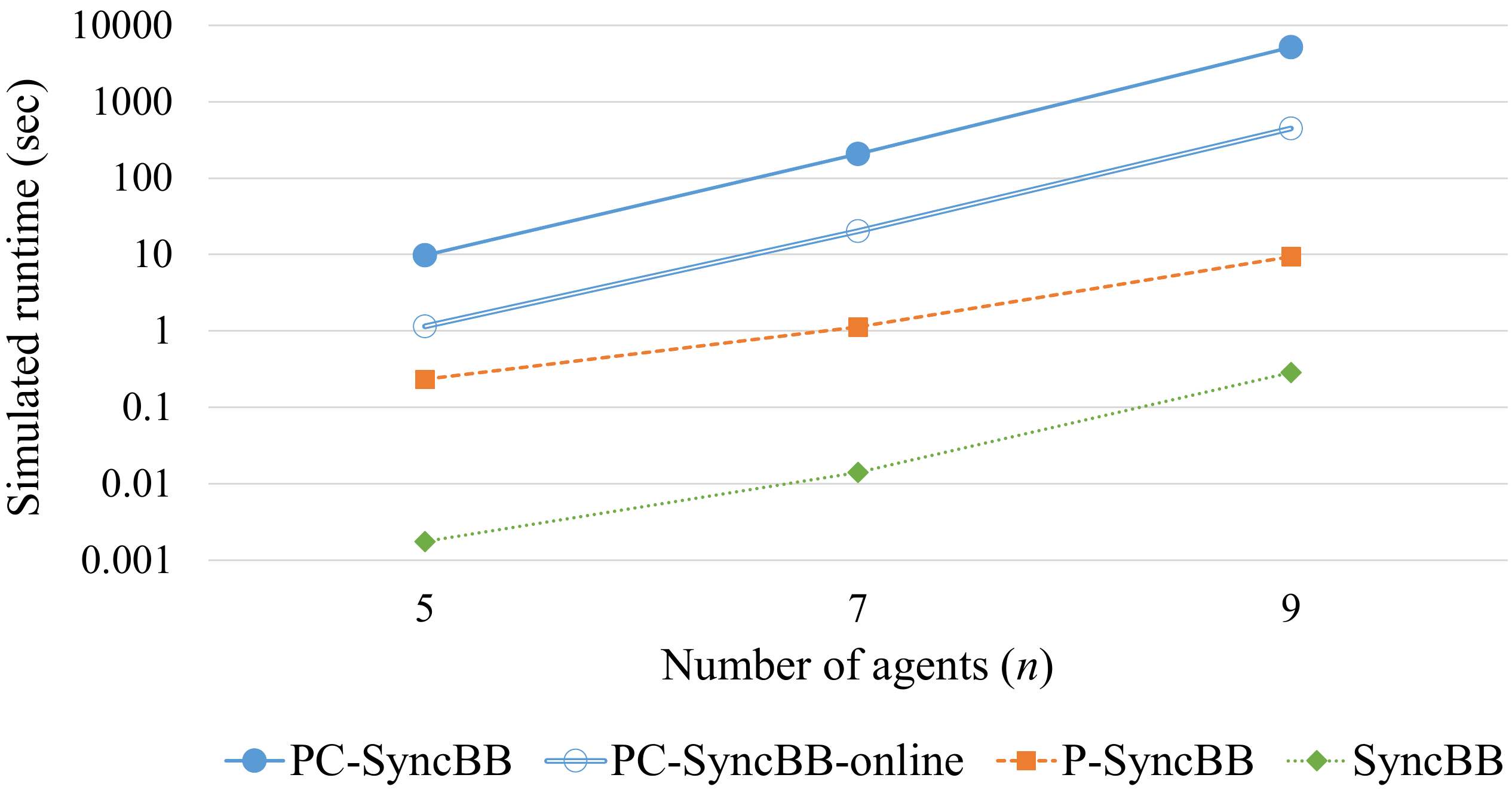}
\vspace{-6pt}
\caption{Varying $n$ (random DCOPs).}
\label{Fig:randomScale}
\end{figure}

In the second experiment, shown in Figure~\ref{Fig:randomScale}, we fix the constraint density to $p_1=0.3$ and the domain sizes to 6, and vary the number of agents $5 \leq n \leq 9$. Here and in the following scalability experiments we use a cutoff time of 30 minutes for online PC-SyncBB. 
It is clear that the number of agents has a major effect on the performance of all the evaluated algorithms, in accordance with known results regarding the scalability of Branch \& Bound algorithms in computationally hard problems. Interestingly, P-SyncBB scales slightly better, probably due to its inherent use of sorted value ordering.

\vspace{-2pt}

\begin{figure}[h!!!]
\centering
\includegraphics[scale=0.25]{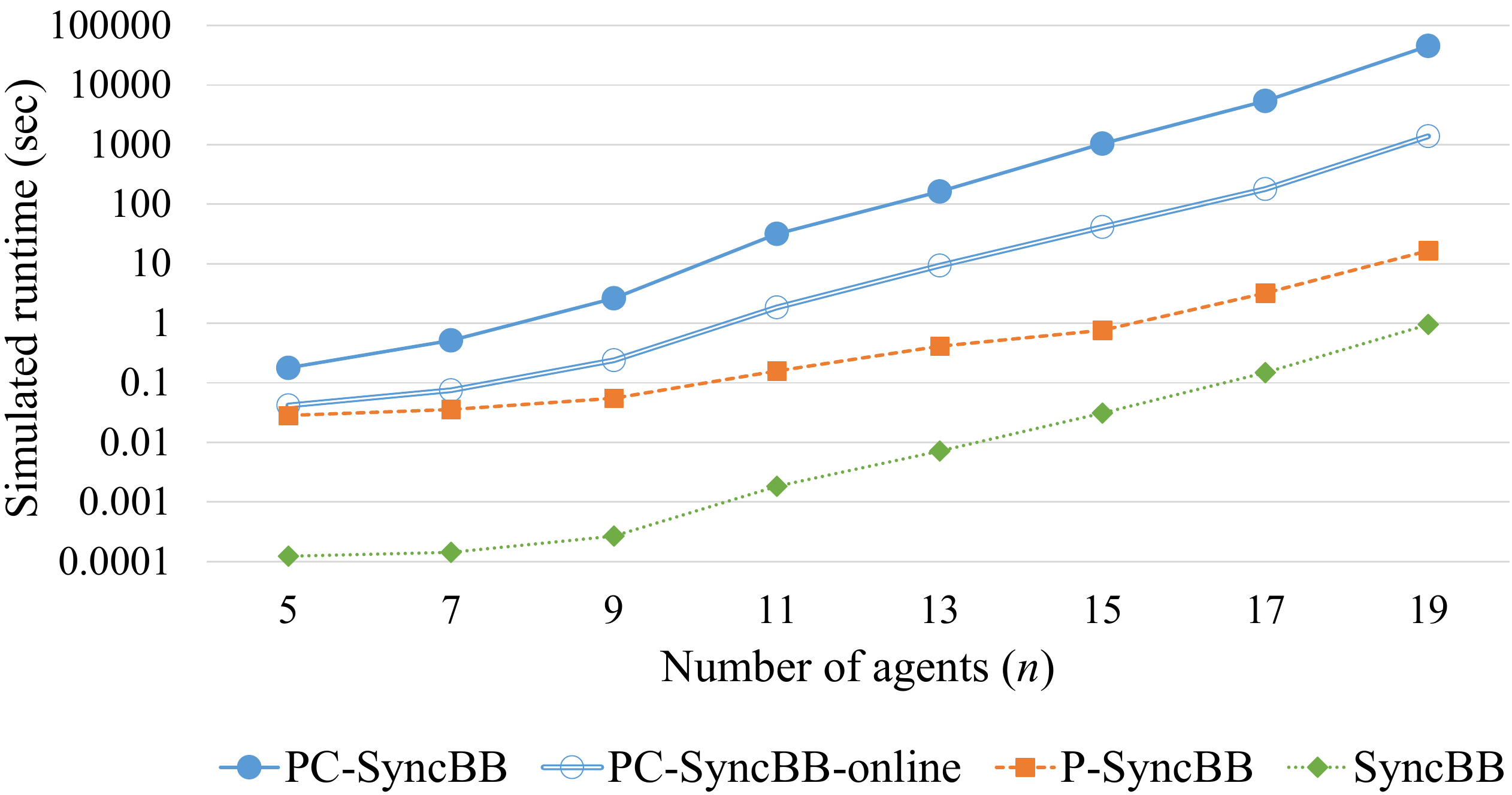}
\vspace{-6pt}
\caption{Varying $n$ (graph coloring problems).}
\label{Fig:gcScale}
\end{figure}

Similar scalability phenomena are also observed in more structured benchmarks. Figure~\ref{Fig:gcScale} depicts the runtime performance on distributed 3-color graph coloring problems ($p_1=0.4$, $5 \leq n \leq 19$) in which each pair of equal values of constrained agents imposes a random and {\em private} cost. The structure in these problems lies in the diagonal constraint matrices between every pair of neighboring agents.

\begin{figure}[h!!!]
\centering
\includegraphics[scale=0.25]{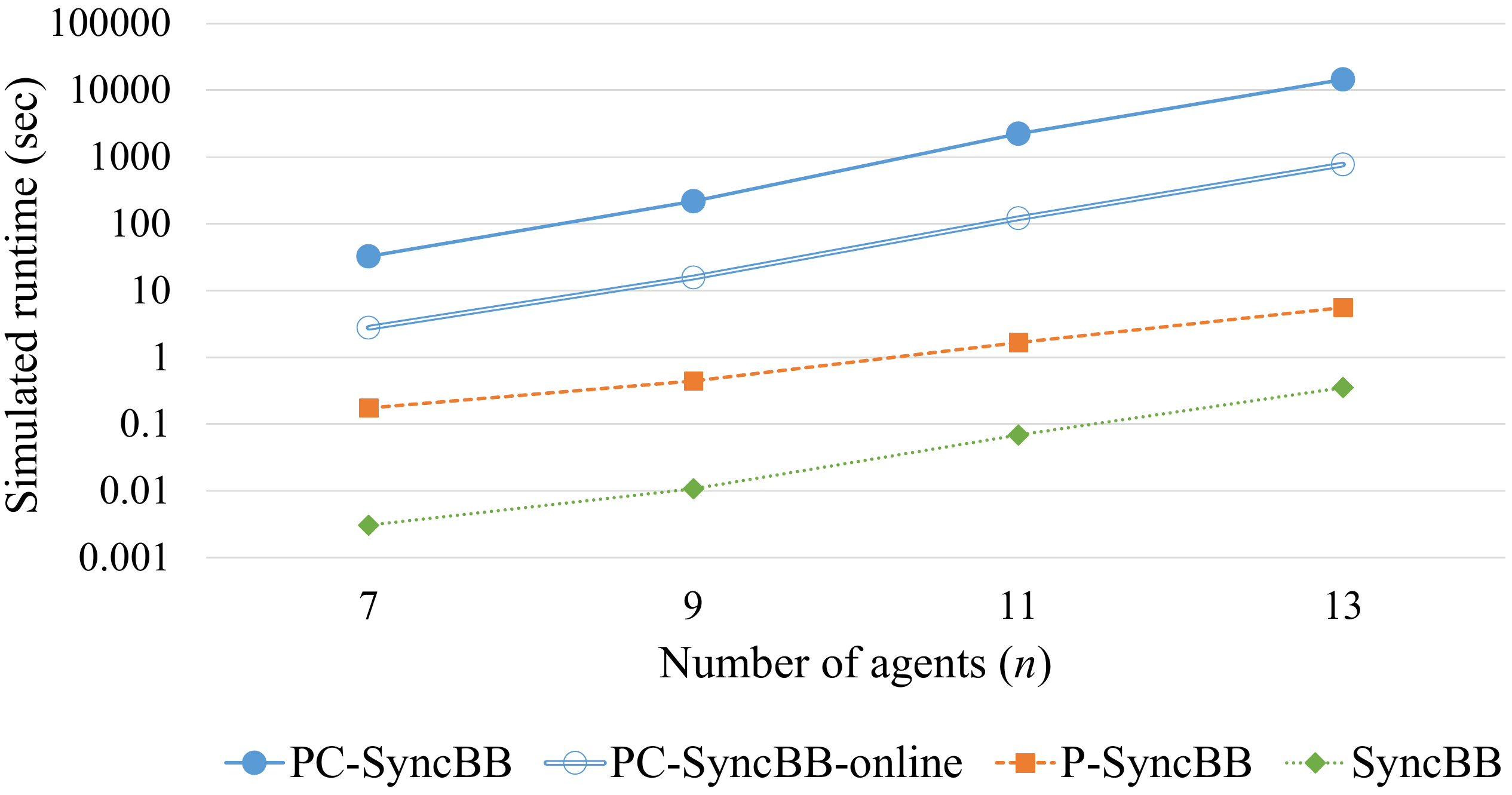}
\vspace{-6pt}
\caption{Varying $n$ (scale-free networks).}
\label{Fig:sfScale}
\end{figure}

Figure~\ref{Fig:sfScale} presents the runtime performance on scale-free networks ($7 \leq n \leq 13$, domains of size 5), which are structured networks that are generated according to the Barab\'{a}si-Albert model \cite{barabasi1999emergence}.

\vspace{-3pt}

\vspace{-3pt}
\vspace{-2pt}

\comment{

Experiments on communication costs show similar trends, but are omitted due to page limitations.

As demonstrated by our experiments, a significant portion of runtime is spent in compare\_CPA\_cost\_to\_upper\_bound.
In our implementation we used a protocol that requires performing $4n^2$ cryptographic operations for every \texttt{AND} gate, which incurs a huge runtime overhead.
There are two promising directions to mitigate this cost. One option is to use the GMW protocol \cite{GMW87}
for a very efficient MPC of Boolean circuits. The GMW protocol requires to perform $O(n^2)$ bit Oblivious Transfers \cite{OT},
but there exist extremely fast Oblivious Transfer protocols \cite{AsharovL0Z15}. Another option is to use a protocol for secure {\em arithmetic} circuit computation, such as the state of the art protocol of Chida et al. \shortcite{CGHIKLN18}, followed by bit-decomposition for extracting the most significant bit.
The cost of the latter operation depends on the bit length of the shares. As the bit lengths in our case are very modest (see the values of $\ell$ in Table \ref{tbl:mpc-results}), and the GMW protocol's complexity is linear in the number of parties rather than quadratic, then that direction also shows a lot of promise.
}

\section{Conclusion}\label{conc}
We proposed herein PC-SyncBB, the first privacy-preserving DCOP algorithm which is secure against coalitions.
We analyzed the properties of the algorithm and evaluated its performance.
Our experiments demonstrate that PC-SyncBB is feasible for moderately-sized problems.

A major limitation on scalability is due to the protocol of Ben-Efraim and Omri \shortcite{BO17} that is invoked by
compare\_CPA\_cost\_to\_upper\_bound. In the future we intend to explore other directions that could yield much more efficient
implementations of that sub-protocol.
If those constructions prove more efficient, we intend to consider more challenging settings, with larger coalitions or malicious agents.
While raising the security bar usually increases runtime and communication costs, it is important to come up with such solutions in order to enlarge and diversify the toolkit available for implementations in various application settings.
Finally, as all existing privacy-preserving DCOP algorithms base their security on assuming solitary conduct of the agents, we view this study as an essential first step towards lifting this potentially harmful assumption in all those algorithms.
In particular, it is necessary to develop privacy-preserving and collision-secure implementations of other DCOP algorithms (e.g., inference-based), and especially of incomplete algorithms that could offer better scalability.

\newpage
\bibliographystyle{named}
\bibliography{SecurityRefs,DCSPRefs}

\end{document}